# Molten salt synthesis of a nanolaminated Sc$_2$SnC MAX Phase

LI You-Bing[12], QIN Yan-Qing[12], CHEN Ke[12], CHEN Lu[12], ZHANG Xiao[12], DING Hao-Ming[12], LI Mian[12], ZHANG Yi-Ming[12], DU Shi-Yu[12], CHAI Zhi-Fang[12], HUANG Qing[12]

(1. Engineering Laboratory of Advanced Energy Materials, Ningbo Institute of Materials Technology and Engineering, Chinese Academy of Sciences, 315201, Ningbo, China; 2. Qianwan Institute of CNiTECH, Zhongchuangyi Road, Hangzhou bay District, Ningbo, Zhejiang, 315336)

**Abstract:** The MAX phases are a family of of ternary layered material with both metal and ceramic properties, and it is also precursor materials for synthesis of two-dimensional MXenes. The theory predicted that there are more than 600 stable ternary layered MAX phases. At present, there are more than 80 kinds of ternary MAX phases synthesized through experiments, and few reports on MAX phases where M is a rare earth element. In this study, a new MAX phase Sc$_2$SnC with rare earth element Sc at the M sites was synthesized through the reaction sintering of Sc, Sn, and C mixtures. Phase composition and microstructure of Sc$_2$SnC were confirmed by X-ray diffraction, scanning electron microscopy and X-ray energy spectrum analysis. And structural stability, mechanical and electronic properties of Sc$_2$SnC was investigated via density functional theory. This study open a door for explore more unknown ternary layered rare earth compounds Re$_{n+1}$SnC$_n$ (Re=Sc, Y, La-Nd, n=1) and corresponding rare earth MXenes.

The MAX phases are a family of nanolayered ternary carbides or nitrides with a hexagonal lattice structure ($P6_3/mmc$), the chemical formula is M$_{n+1}$AX$_n$ (where M is an early transition metal; A belongs to group of 13-16; and X is carbon or/and nitrogen, n=1-3)[1-3]. Generally, the heterodesmic feature of MAX phases contributes to a unique combination of both metallic and ceramic properties, and have been investigated as promising candidates for structural applications in many fields[4-7]. Moreover, MAX phases are used as a precursor to synthesize two-dimensional (2D) MXene with many attractive physical and chemical properties, and show promise in a broad range of applications, notably electrochemical energy storage[8-11]. Due to the continuing efforts from the scientific community, about 155 MAX phases have been reported so far; including some novel MAX phases that A-site elements are late transition metals[12-16]. The theoretical studies have predicted around 665 ternary MAX phases that could be experimentally synthesized[17], for example, the ones whose M site element is rare earth Sc.

As previous reported, Sc$_2$InC was listed as one of possible stable MAX phases[1,3], where the structure, properties and potential applications are investigated via theoretical predictions[18-20]; but not been experimentally identified yet. The Sc$_2$InC is expected to be a promising candidate for optoelectronic devices for the visible light and ultraviolet regions, as well as coating materials to avoid solar heating[20]. In addition, theoretical calculations indicate that the Sc$_2$CT$_2$ (T= F, OH) MXenes can be promising candidate materials for the next generation electronic devices[21]. S. Kuchida et al[22] focused on non-transition metal M$_2$AX compounds which embody Sc, Y, and Lu atoms in M site; however, only polycrystalline sample of Lu$_2$SnC was reported. As a result, the study of new MAX phases taking Sc as M site element is an intriguing and challenging work.

Now, the common synthesized of MAX phases methods are hot pressing (HP) and spark plasma sintering (SPS). Compared to HP and SPS, the molten salt method is a simple and cost-effective route for preparing MAX phase powders. As a high-temperature ionic solvent, the molten salt bath offers high solvation power and liquid environment for reactants that will greatly facilitate the mass transport and nucleation processes, thus need lower synthesis temperature and bold time[23]. Some MAX phases (e.g. Ti$_3$SiC$_2$, Ti$_3$AlC$_2$, V$_2$AlC, Cr$_2$AlC) have been synthesized by molten salt method[24-28]. In the present work, we synthesized a MAX phase of Sc$_2$SnC in molten salts environment where the Sc element belongs to rare earth. The crystal structure and chemical composition were confirmed by XRD and SEM-EDS, respectively. Furthermore, the structure stability, electronic structure and mechanical properties of Sc$_2$SnC are also be investigated via density functional theory (DFT).

# 1 Experimental

The raw materials used to prepare the MAX phase are scandium (Hunan Rare Earth Metal Materials Research Institute, Hunan, China; ~300 mesh, 99.5 wt.% purity), tin (Target Research Center of General Research Institute for Nonferrous Metals, Beijing, China, ~300 mesh, 99.5 wt.% purity), graphite (Qingdao Tianshengda Graphite Co. Ltd, Shandong, China; ~300 mesh, 99 wt.% purity), sodium chloride (Aladdin Industrial Co. Ltd, Shanghai, China; NaCl, 99.5 wt.% purity), potassium chloride (Aladdin Industrial Co. Ltd, Shanghai, China; KCl, 99.5 wt.% purity).

The powders were mixed in a stoichiometric ratio of Sc: Sn: C = 2: 1.1: 1 (Due to the melting point of Sn is relatively low, we increased the content ratio of tin for compensating the weight loss of tin at a high temperature, as in the preparation of $V_2(Sn,A)C$ MAX phases)[16]. The starting powders of Sc, Sn and graphite are mixed with inorganic salt (NaCl + KCl), and the mole ratio of (Sc+Sn+C):(NaCl+KCl) was 1:10. After ground for 10 min, the powder mixture was put into an aluminum oxide boat, and then move to a tube furnace and heated to 1000 °C during 3 h with heating rate of 5 °C/min under argon atmosphere, respectively. After the reaction was finished, the product is washed, filtered and dried at 40 °C in vacuum; and the excess Sn element is removed by ferric chloride (Aladdim Industrial Co. Ltd, Shanghai, China; $FeCl_3$, 99.5 wt.% purity).

The phase composition of the samples was determined by X-ray diffraction (XRD, D8 Advance, Bruker AXS, Germany) with Cu $K_\alpha$ radiation. X-ray diffraction patterns were collected at a step size of 0.02° 2θ with a collection time of 1 s per step. The microstructure and chemical composition were observed by scanning electron microscopy (SEM, QUANTA 250 FEG, FEI, USA) equipped with an energy-dispersive spectrometer (EDS), and the EDS values were fitted by XPP (extended Puchou/Pichoir).

Density functional theory (DFT) calculations were programmed in the CASTEP code[29-30], using the generalized gradient approximation (GGA) as implemented in the Perdew-Breke-Ernzerhof (PBE) functional[31-32]. Phonon calculations were carried out to evaluate the dynamical stability using the finite displacement approach, as implemented in CASTEP[33-34]. The equation $E=(E_{broken}-E_{bulk})/S$[13] was adopted to calculate the cleavage energy $E$, where $E_{bulk}$ and $E_{broken}$ represent the total energies of bulk MAX and the cleaving structures respectively with a 10 Å vacuum separation in the corresponding M and A atomic layers, and $S$ is the cross-sectional surface area of the MAX phase materials. The Rietveld refinement of powder XRD pattern of $Sc_2SnC$ was by TOtal PAttern Solution (TOPAS-Academic V6) software.

# 2 Results and discussion

## 2.1 Phase analysis of the $Sc_2SnC$

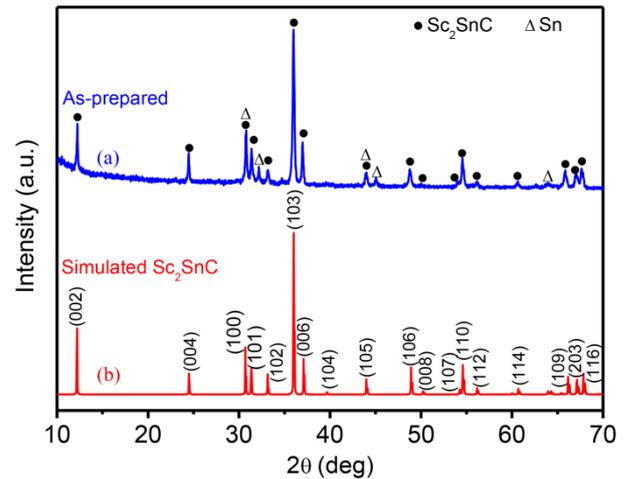

Fig. 1. Comparison of XRD patterns between (a) powders synthesized through the reaction between Sc, Sn, and C mixtures and (b) the simulated one of $Sc_2SnC$

Fig. 1a shows the XRD pattern of as-prepared powders synthesized at 1000 °C for 3 h, with characteristic peaks at 2θ ~ 12°, 2θ ~ 24°, and 2θ ~ 36°, which neither belong to Sn nor other compounds, indicating a new MAX phase of $Sc_2SnC$ was synthesized (also with minor amount of Sn metal as by-product). In comparison with experimental result, the simulated XRD pattern of $Sc_2SnC$ in Fig. 1b, the peaks positioned at 12.174°, 24.517°, 36.277°, etc., consisting with the experimental peak positions in Fig. 1a, which further validates the formation of the new MAX phase $Sc_2SnC$.

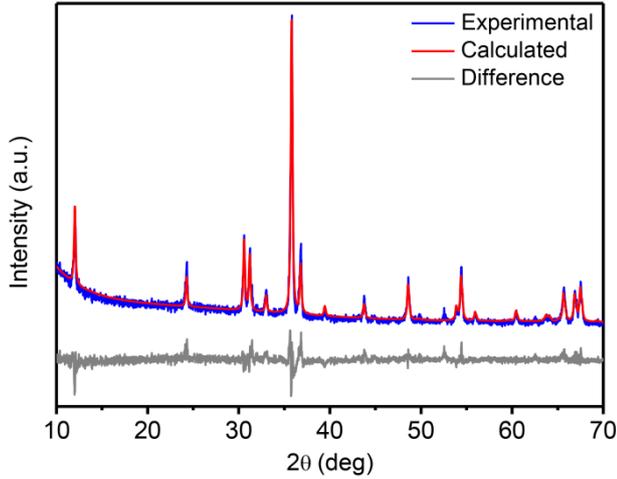

Fig. 2. Comparison between experimental (blue crosses) and calculated XRD (red line) pattern of $Sc_2SnC$.

Table 1. Atomic positions in $Sc_2SnC$ determined from the Rietveld refinement

| Site | Element | $x$ | $y$ | $z$ | Symmetry | Wyckoff symbol |
|------|---------|-----|-----|-----|----------|----------------|
| M | Sc | 1/3 | 2/3 | 0.5786 | 3m | 4f |
| A | Sn | 1/3 | 2/3 | 0.25 | $\bar{6}m2$ | 2d |
| X | C | 0 | 0 | 0 | $\bar{3}m$ | 2a |

XRD pattern is important for phase identification and structure analysis. Due to no XRD pattern of $Sc_2SnC$ is available from literature, the Rietveld refinement of powder XRD pattern of $Sc_2SnC$ was conducted. As shown in Fig. 2, the blue crosses represent the experimental diffraction profile (the Sn metal was remove by $FeCl_3$ solution), while the red solid line denotes the theoretical pattern. The theoretical Bragg diffraction positions of $Sc_2SnC$ are marked as green lines. The gray curve is the deviation between calculated and experimental XRD patterns. The obtained reliability factors are $R_p$ = 8.56% and $R_{wp}$ = 11.19%, respectively; indicating good agreement between model and measured data. The space group of $Sc_2SnC$ is $P6_3/mmc$ (No. 194), and the lattice constants measured from XRD pattern are $a$ = 3.3692 Å and $c$ = 14.6374 Å, respectively. The difference between theoretical calculation and the Rietveld refinement is probably ascribed to the existence of defects in the crystal structure, as the case of $V_2SnC$ in previous report[35]. The atomic positions of $Sc_2SnC$ determined from the Rietveld refinement are listed in Table 1.

## 2.2 Microstructural of the $Sc_2SnC$

It is well known that MAX phases crystallize in hexagonal structures and their grains are generally layered hexagonsin morphology[1]. To confirm that $Sc_2SnC$ has a similar microstructure, the microstructure of as-prepared powder was observed by SEM. It can be seen from Fig. 3a that $Sc_2SnC$ exhibits the microstructure of typical thin hexagons. EDS equipped in SEM detected all constitutive elements (Sc, Sn and C) within these particles (shown in Fig. 3b). Although the EDS analysis is semi-quantitative and the accurate determination of light elements like C is difficult, the relative atomic ratio of (Sc:Sn:C) could be revealed by EDS as about (2:1:1), consistent with the stoichiometry of 211 MAX phases. The Elemental mapping of Sc, Sn and C corroborated that all of these three elements have the same distribution. The above results further confirmed that the new MAX phase compound $Sc_2SnC$ is experimentally synthesized.

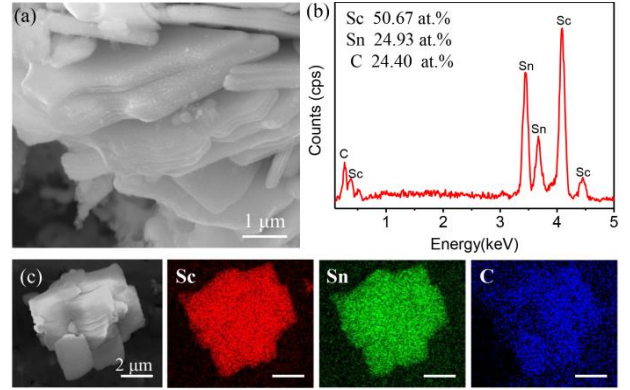

Fig. 3. (a) SEM image of $Sc_2SnC$. (b) The corresponding EDS spectrum indicated that particles contain Sc, Sn and C elements. (c) Elemental mapping clearing proving the uniform distribution

## 2.3 DFT results

The structural analysis of $Sc_2SnC$ phase was carried out via DFT calculations. Fig. 4a shows the ternary-layered carbide crystal structure of $Sc_2SnC$; and the calculated $\Delta H_{form}$ ($Sc_2SnC$) is –0.7167 eV, indicating stability of the $Sc_2SnC$ phase. The lattice parameters, elastic constants and polycrystalline elastic modulus of $Sc_2SnC$, as well as for other Sn-containing MAX phases are listed in Table 2. From the DFT calculation result, the $a$= 3.3686, $c$= 14.6532 Å, which was very close to experimental results. The mechanical stability of $Sc_2SnC$ was justified from the Born stability criteria[36]: $C_{11} > 0$, $C_{11}-C_{12} > 0$, $C_{44} > 0$, $(C_{11}-C_{12})C_{33}-2C_{13} > 0$. Besides, the dynamical stability of $Sc_2SnC$ can also be identified from the phonon dispersion curves in Fig. 4b. The results performed by theoretical calculations are consistent with experiments. However, compared with other MAX phases (listed in Table 2), it is found that they have lower values of elastic constants (i.e. $C_{11}$, $C_{33}$, $C_{44}$, and $C_{66}$).

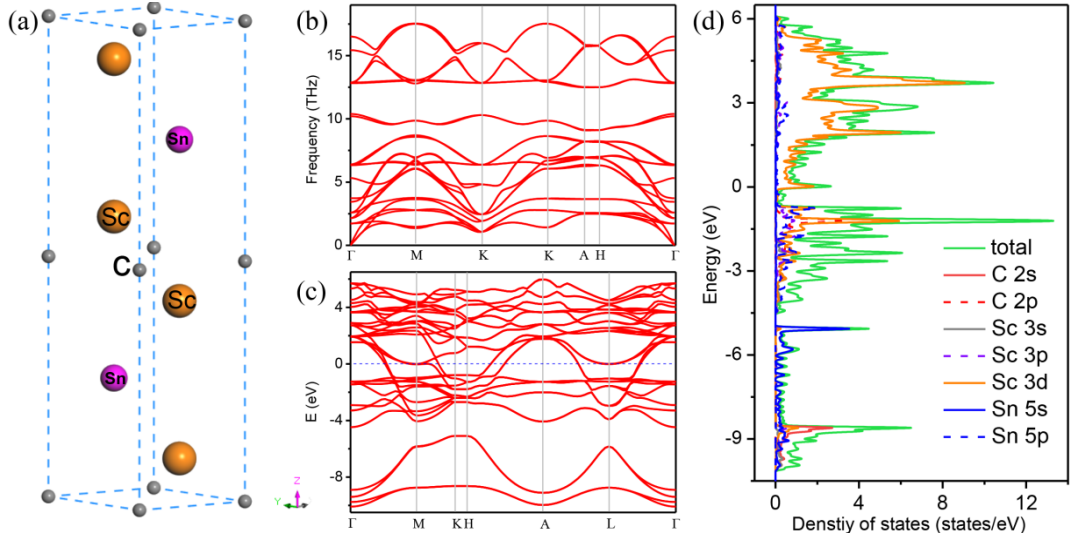

Fig. 4. (a) Crystal structure of Sc$_2$SnC; (b) Calculated phonon dispersion of Sc$_2$SnC; (c) Band structure of Sc$_2$SnC; (d) Projected density of Sc, Sn, and C atom states.

Table 2. Theoretically predicted Lattice parameters (Å), calculated elastic constants, $C_{ij}$ (GPa), bulk modulus, B (GPa), shear modulus, G (GPa), and Young's modulus, E (GPa), Pugh ratio, G/B, and Poisson ratio, $v$, of Sc$_2$SnC, V$_2$SnC, Ti$_2$SnC, Zr$_2$SnC, Hf$_2$SnC, and Nb$_2$SnC.

| Compound | $a$ (Å) | $c$ (Å) | $C_{11}$ | $C_{12}$ | $C_{13}$ | $C_{33}$ | $C_{44}$ | $C_{66}$ | B | G | E | G/B | $v$ | Ref. |
|---|---|---|---|---|---|---|---|---|---|---|---|---|---|---|
| Sc$_2$SnC | 3.368 | 14.653 | 197 | 63 | 47 | 182 | 67 | 53 | 100 | 63 | 157 | 0.630 | 0.238 | This work |
| V$_2$SnC | 3.134 | 12.943 | 336 | 126 | 122 | 304 | 85 | 105 | 190 | 95 | 244 | 0.500 | 0.286 | 35 |
| Ti$_2$SnC | 3.136 | 13.641 | 337 | 86 | 102 | 329 | 169 | 126 | 176 | 138 | 328 | 0.784 | 0.188 | 39 |
| Zr$_2$SnC | 3.352 | 14.681 | 269 | 80 | 107 | 290 | 148 | 94 | 157 | 110 | 368 | 0.700 | 0.215 | 39 |
| Hf$_2$SnC | 3.308 | 14.450 | 330 | 54 | 126 | 292 | 167 | 138 | 173 | 132 | 316 | 0.763 | 0.195 | 39 |
| Nb$_2$SnC | 3.244 | 13.754 | 341 | 106 | 169 | 321 | 183 | 118 | 209 | 126 | 314 | 0.603 | 0.250 | 39 |

The relative small value of $C_{33}$ indicates the compound is more compressible along the c-axis compared to other compounds studied; while low $C_{44}$ indicates being subject to shear deformation along [11$\bar{2}$0] (0001); and small $C_{66}$ probably means lower resistance to shear in the <110> direction[20,35,37]. The low shear deformation of Sc$_2$SnC is also reflected from shear modulus G, which represents the resistance to shape change of the polycrystalline material[38]. The calculated value of G/B > 0.5 indicates that the phase is brittle in nature following Pugh's criterion. Furthermore, the obtained value of $v$ (0.238) for Sc$_2$SnC shows that it is located at the boundary between covalent and ionic materials. The calculated band structure of Sc$_2$SnC and the projected density of states (DOS) of Sc, Sn, and C atoms with k-points are shown in Fig. 4c and Fig. 4d, respectively. Similar to other MAX phases and MAX phase-like compounds, Sc$_2$SnC exhibits metallic nature; and the overlapping between valence and conduction bands across the Fermi level also reveals the presence of metallic bonding, which can be treated as the origin of the quasi-ductility of Sc$_2$SnC (Fig. 4c). As can be seen from Fig. 4d, it is can be observed that the Sc-3$d$ electrons are mainly contributing to the DOS at the Fermi level, and should be involved in the conduction properties; while the contributions from.

# 3 Conclusions

In conclusion, a new MAX phase was successfully synthesized for the first time by heating the Sc, Sn, and C raw powder mixtures. The XRD data of $Sc_2SnC$ is useful for further phase identification and structure analysis; and $Sc_2SnC$ exhibits a typical laminar microstructure similar to other MAX phases. The first-principle calculations were employed to further study structure stability of $Sc_2SnC$ MAX phase, and the results showed that $Sc_2SnC$ is metallic in nature where the contribution from Sc-$3d$ states dominates the electronic conductivity at the Fermi level. This work implies the great potential of the ternary rare-earth metal carbide $Re_{n+1}SnC_n$ (Re=Sc, Y, La-Nd, n = 1) family waiting for further explore. More importantly, the introduction of rare earth elements can give the special properties of MAX phases, and can be used as a precursor material for preparing rare-earth MXenes.